\documentclass{llncs}
\usepackage{epsfig}
\usepackage{amsmath}
\usepackage{amssymb}

\begin{document}

\title{CoPhIR: a Test Collection for Content-Based Image Retrieval}


\author{Paolo Bolettieri \and Andrea Esuli \and Fabrizio Falchi \and
  Claudio Lucchese \and \\ Raffaele Perego \and Tommaso Piccioli \and
  Fausto Rabitti} \institute{ISTI-CNR, Pisa, Italy.\\
  \email{firstname.lastname@isti.cnr.it}}

\maketitle

\begin{abstract}
The scalability, as well as the effectiveness, of the different
Content-based Image Retrieval (CBIR) approaches proposed in
literature, is today an important research issue. Given the
wealth of images on the Web, CBIR systems must  in fact leap towards
Web-scale datasets. In this paper, we report on our experience in
building a test collection of 100~million images, with the corresponding descriptive
features, to be used in experimenting new scalable  techniques for
similarity searching, and comparing their results. In the context
of the SAPIR (Search on Audio-visual content using Peer-to-peer
Information Retrieval) European project, we had to experiment our
distributed similarity searching technology on a realistic data
set. Therefore, since no large-scale collection was available
for research purpose, we had to tackle the non-trivial process of
image crawling and descriptive feature extraction (we used five
MPEG-7 features) using the European EGEE computer GRID. The result
of this effort is CoPhIR, the first CBIR test collection of such scale. CoPhIR
is now open to the research community for experiments and
comparisons, and access to the collection was already granted to more than $50$ research groups worldwide.

\end{abstract}

\section{Introduction}
\label{sec:intro}

Everybody knows about the data explosion. According to recent
studies, in the next three years, we will create more data than
has been produced in all of human history. Regarding images, the
Enterprise Strategy
Group\footnote{http://www.enterprisestrategygroup.com/} estimates
that more than 80 billion photographs are taken each year.
Storing them would require 400 petabytes of storage. Therefore the
management of digital images promises to emerge as a major issue
in many areas providing a lot of opportunities in the next years,
particularly since a large portion of pictures still remains as
``unstructured data'', i.e., with no meaningful associated tags.

Current searching engines headed by Google are in the center of
current information age; Google answers daily more
than 200 million queries against over 30 billion items.
However, the search power of these engines is typically limited to
text and its similarity. Since less than 1\,\% of the Web data
is in textual form, the rest being of multimedia/streaming
nature, we need to extend our next-generation search to
accommodate these heterogeneous media. Some of the current engines
search these data types according to textual information or other
attributes associated with the files.

An orthogonal approach is the Content-based Image Retrieval
(CBIR). It is not a new area as demonstrated by a recent survey
\cite{Datta2008} which reports on nearly 300 systems, most of them
exemplified by prototype implementations.  However, the typical
database size is in the order of thousands of images. Very recent
publicly-available systems, such as
ImBrowse\footnote{http://media-vibrance.itn.liu.se/},
Tiltomo\footnote{http://www.tiltomo.com/} or
Alipr\footnote{http://www.alipr.com/}, declare to index hundreds
of thousands of images. There is a high discrepancy between these
numbers and the volumes of images available on current Web, so we
decided to investigate the situation by shifting the current
bounds up by two orders of magnitude.

This work has been developed within the European project SAPIR
(Search on Audio-visual content using Peer-to-peer Information
Retrieval)\footnote{SAPIR European Project, IST FP6:
http://www.sapir.eu/}. This project aims at finding new
content-based methods to analyze, index, and retrieve the
tremendous amounts of speech, image, video, and music which are
filling our digital universe. In this context, we intended to
develop a large-scale distributed architecture for indexing and
searching in image collections according to visual characteristics
of their content. The system should be able to scale to the order
of tens of millions. To reach this goal, a collection of such size
together with respective descriptive features is needed. We have
crawled the public images from a popular photo-sharing system
Flickr\footnote{http://www.flickr.com/} and have extracted five
MPEG-7 features from every image. This source has also the
advantage of having associated user-defined textual information,
which could be used for experiments combining search
on text and visual content.

Our scalability objective, $100$ million images, goes very far beyond the current
practice. On the data acquisition level, this would require to
download and process 30\,TB to 50\,TB of data, depending on the
image resolution. Moreover, we need a storage space for the image
descriptors (including the MPEG-7 features). In practice,  we have to bear in mind that the
image crawling and feature extraction process would take about 12
years on a standard PC and about 2 years using a high-end
multi-core PC.

The rest of the paper is organized as follows. Section
\ref{sec:crawl} describes the process of building the test
collection: crawling the images, extracting the MPEG-7 features,
and organizing the result into the test collection which is now
open to research community. Finally, Section \ref{sec:concl}
analyzes the results obtained.

\section{Building the Image Collection}
\label{sec:crawl}
Collecting a large amount of images for investigating CBIR issues is
not an easy task, at least from a technological point of view. The
challenge is mainly related to the size of the collection we are
interested in.  Shifting state-of-the-art bounds of two orders of
magnitude means building a $100$ million collection, and this size
makes very complex to manage every practical aspect of the gathering
process. However, since the absence of a publicly available collection
of this kind has probably limited the academic research in this
interesting field, we tried to do our best to overcome these
problems. The main issues we had to face were:

\begin{enumerate}
 \item identification of a valid source;
 \item efficient downloading and storing of such a large collection of images;
 \item efficient extraction of metadata (MPEG-7 visual descriptors and others) from the downloaded images;
 \item providing reliable data-access to metadata.
\end{enumerate}

In the following we will discuss the above issues by describing the
challenges, the problems we encountered, and the decisions we took.

\subsection{Choosing the Data Source}
Crawling the Web is the first solution if you are looking for a
practically unlimited source of data. There are plenty of images on
the Web, varying in quality from almost professional to amateur
photos, from simple drawings to digital cartoons.

There are also many different ways to retrieve such data. The
first option is to exploit spider agents that crawl the Web and
download every image found on the way. Of course this would result
in a large amount of time and bandwidth wasted in downloading and
parsing HTML pages, possibly gathering only a few images. The
authors of \cite{imagesperpage} report that the average number of
images hyperlinked by HTML pages is varying. In their experiments
with the Chilean Web, they repeatedly downloaded each time about
1.3 million Web pages.  The number of images retrieved were
100,000 in May 2003, 83,000 in August 2003 and 200,000 in January
2004. Thus, assuming that these percentages are still valid today,
we can expect that to gather 100 million images, we would have to
download and parse 650 million to 1.5 billion Web pages.

A second option, which may be more efficient, is to take advantage
of the image search service available on most commercial Web
search engines. Just feeding the search engine with queries
generated synthetically, or taken from some real query log, would
provide us with plenty of images.

This abundance and diversity of Web images is definitely a plus.
Not only because we want a large collection, but also because we
want our collection to spread over different kinds of images. A
problem is instead given by the large differences in the quality
and size of the retrieved images. A large portion of them are
decoration elements like buttons, bullet list icons, and many
other are very small images or photo thumbnails. These images are
not suitable for our purposes and would pollute the corpus,
but some of them could be filtered out automatically as the
feature extraction software is likely to fail on images
with non-standard sizes.

However, for the need of high-quality data, we finally decided to
follow a third way: crawling one of the popular photo sharing
sites born in the last years with the goal of providing permanent
and centralized access to user-provided photos. This approach has
several advantages over the aforementioned approaches.

\paragraph*{Image Quality}
In fact photo sharing sites like Flickr, PhotoBucket, Picasa,
Kodak EasyShare Gallery, Snapfish, etc.  mainly store high-quality
 photographic images. Most of them are very large since they come from
 3--8 Megapixel cameras, and have a standard 4:3 format.

\paragraph*{Collection Stability}
These sites provide quite static, long term and reliable
image repositories. Although images may be deleted or made private
by the owners, this happens quite rarely. Most photos stay
available for a long time and they are always easily accessible.
Conversely, the Web is much more dynamic, images change or are
moved somewhere else, pages are deleted and so on.

\paragraph*{Legal Issues}
The above consideration is very important also when considering
the legal issues involved in the creation of such collection of
images. In fact, storing for a long time a publicly available image
may in some case violate author's copyrights. We are mainly interested
in the visual descriptors extracted from the images, but any
application of CBIR has to access the original files for eventually
presenting the results retrieved to a human user. Since Photo sharing
sites are fairly static, we can build a quite stable collection
without permanently storing the original files, but maintaining only
the hyperlinks to the original photos that can be accessed directly at
any time.

\paragraph*{Rich Metadata}
Finally, photo sharing sites provide a significant
amount of additional metadata about the photos hosted. The digital
photo file contains information about the camera used
to take the picture, the time when it was taken, the aperture,
the shutter used, etc. More importantly, each photo comes with the
name of the author, its title, a description, often with user-provided
tags. Sometimes also richer information is available such as comments
of other users on the photo, the GPS coordinates of the location where
the photo was taken, the number of times it was viewed, etc.


\medskip

Among the most popular photo sharing sites, we chose to crawl
Flickr, since it is one with the richest
additional metadata and provides an efficient API\footnote{
http://www.flickr.com/services/api/} to access its content at various
levels.

\subsection{Crawling the Flickr Contents}
\label{subsec:crawl}
It is well known that the graph of Flickr users, similarly to all
other social media applications, is scale free~\cite{evolution}. We thus
exploited the small-world property of this kind of graphs to build our
huge photo collection.  By starting from a single Flickr user and
following friendship relations, we first downloaded a partial snapshot
of the Flickr graph. This snapshot of about one million distinct users
was crawled in February 2007. We then exploited the Flickr API to get
the whole list of public photo IDs owned by each of these
users. Since Flickr Photo IDs are unique and can be used to
unequivocally devise an URL accessing the associated photo, in this
way we have easily created a 4.5\,GB file with 300 million
distinct photo IDs.

In the next step, we decided what information to download
for each photo.  Since the purpose of the collection is to enable
a general experimentation on various CBIR research solutions, we
decided to retrieve almost all information available. Thus,
for each photo: title and description, identification and location of the
author, user-provided tags, comments of other users, GPS
coordinates, notes related to portions of the photo, number of
times it was viewed, number of users who added the photo to
their favourites, upload date, and, finally, all the information
stored in the EXIF header of the image file. Naturally, not all
these metadata are available for all photos.  In
order to support content based search, we extracted several
MPEG-7 {\em visual descriptors} from each image~\cite{mpeg7}. A
visual descriptor characterizes a particular visual aspect of the
image. They can be, therefore, used to identify images which have
a similar appearance. Visual descriptors are represented as
vectors, and the MPEG-7 group proposed a distance measure for each
descriptor to evaluate the similarity of two
objects~\cite{mpeg7-2002}. Finally, we have chosen the five MPEG-7
visual descriptors described below~\cite{mpeg7v2002,mpeg7xm}:

\begin{description}
\item [Scalable Colour] It is derived from a colour histogram
defined in the Hue-Saturation-Value colour space with fixed colour
space quantization. The histogram values are extracted, normalized
and nonlinearly mapped into a four-bit integer representation.
Then the Haar transform is applied. We use the 64 coefficients
version of this descriptor.
\item [Colour Structure] It is also based on colour histograms but aims
at identifying localized colour distributions using a small structuring
window. We use the 64 coefficients version of this descriptor.
\item [Colour Layout] It is obtained by applying the DCT transformation
on a 2-D array of local representative colours in Y or Cb or Cr colour
space. This descriptor captures both colour and spatial
information. We use the 12 coefficients version of this descriptor.
\item [Edge Histogram] It represents local-edge distribution in the
image. The image is subdivided into $4\times 4$ sub-images, edges in each
sub-image are categorized into five types: vertical, horizontal,
$45^\circ$ diagonal, $135^\circ$ diagonal and non-directional
edges. These are then transformed in a vector of 80 coefficients.
\item [Homogeneous Texture] It characterizes the region texture using
the mean energy and the energy deviation from a set of 30 frequency
channels. We use the complete form of this descriptors which consist of 62 coefficients.
\end{description}

There are several other visual descriptors which can be useful,
for example, for specialized collections of images (e.g. medical).
Many experiences
suggest that retrieval based on these five MPEG-7 standard
descriptors can be acceptable on non-specialized images, such as
the ones in our collection.

Unfortunately, the extraction of MPEG-7 visual descriptors from
high-quality images is very computationally expensive. Although
the MPEG-7 standard exists for many years, there is not an
optimized extraction software publicly available. To extract
descriptors, we used the MPEG-7 eXperimentation Model (MPEG-7 XM)
\cite{mpeg7xm} that is the official software certified by the MPEG
group that guarantees the correctness of the extracted features.
This software running on a AMD Athlon XP 2000+ box takes about 4
seconds to extract the above five features from an image of size
$500\times 333$ pixels. Therefore, even without considering the
time needed to download the image and all additional network
latencies involved, we can estimate that a single standard PC
would need about $12$ years to process a collection of $100$ million
images.

\begin{figure}
\centering
\includegraphics[width=8cm]{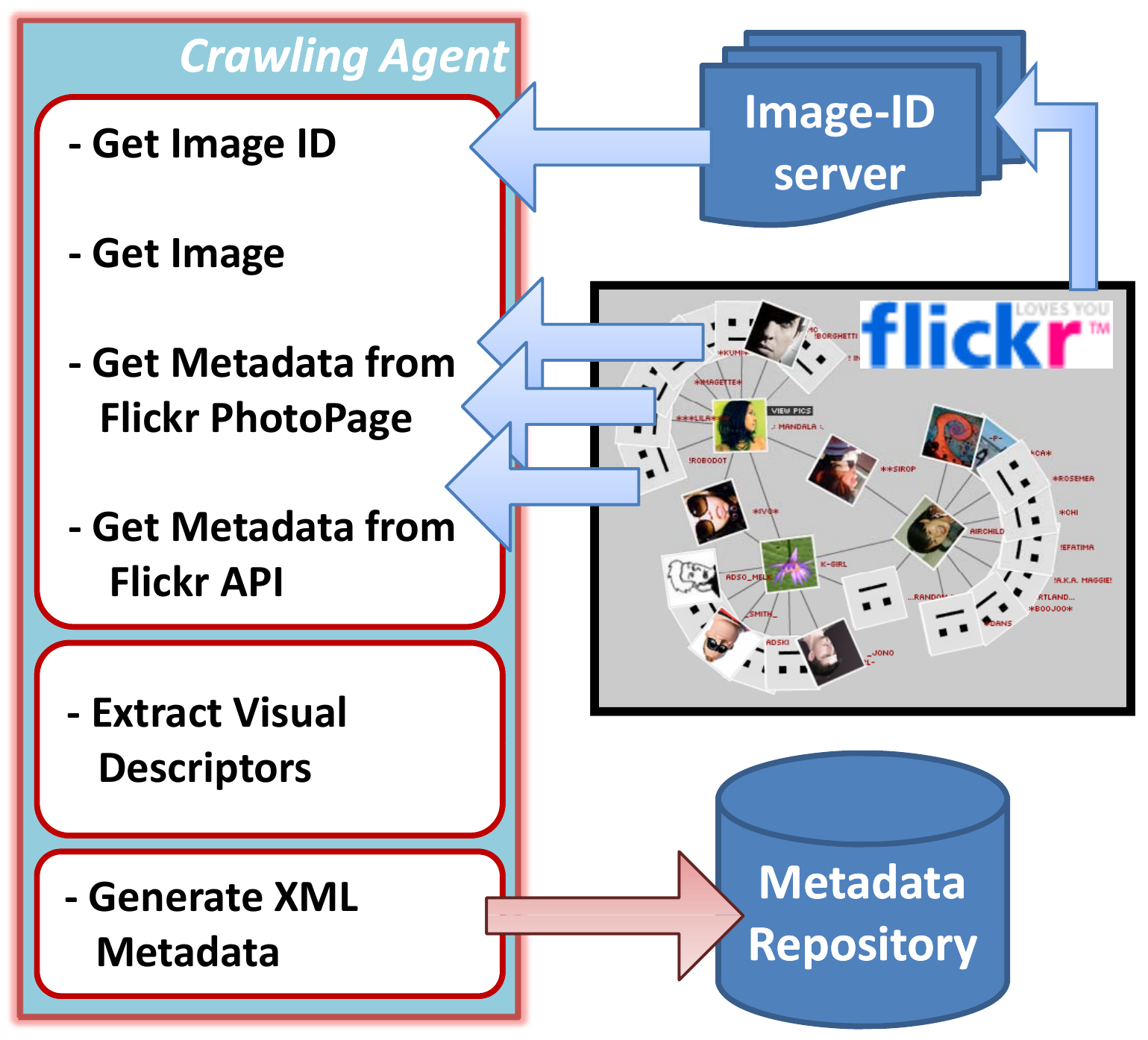}
\caption{\label{fig:crawler}Organization of the crawling and feature extraction process.}
\end{figure}

It was thus clear that we needed a large number of machines working in
parallel to achieve our target collection of 100 million images in a
reasonable amount of time. For this reason, we developed an application
that allows to process images in parallel on an arbitrary (and
dynamic) set of machines. This application is composed of three main
components: the {\em image-id server}, the {\em crawling agents}, and
the {\em repository manager} as shown in Figure~\ref{fig:crawler}.

\medskip
\noindent
 {\bf The image-id server} was implemented in PHP as a simple
Web application accomplishing the task of providing
crawling agents with an arbitrary number of photo identifiers not yet
processed.


\medskip
\noindent
 {\bf The crawling agent} is the core part of our application. It
loops asking the image-id server for a new set of image identifiers to
process. Once it obtains a set from the server, it starts the actual
retrieval and feature extraction process. Given a photo ID, the first
step is to issue an HTTP request and download the corresponding {\em
Flickr photo-page}. This is parsed to retrieve the URL of the image
file and some of the metadata discussed above. Thanks to Flickr APIs,
this metadata is then enriched with other information (title of
the photo, description, tags, comments, notes, upload date,
user name, user location, GPS coordinates, etc.).

We downloaded medium-resolution version of the photos, which have
the larger dimension 500 pixels. This improves the
independence of extracted features from image size and reduces
the cost of processing large images. The MPEG-7 XM~\cite{mpeg7xm} software is used
to extract the aforementioned five visual descriptors.

The extracted features and all the available metadata are used to
produce an XML file containing the knowledge we collected about the
image. Finally, a thumbnail is also generated from the photo.  The XML
file and the thumbnail of the image are sent to a Web-service provided
by the {\em repository manager}.

\medskip
\noindent {\bf The repository manager} runs on a large file-server
machine providing 10\,TB of reliable RAID storage.  In addition to
receive and store the results processed by the crawling agents,
the repository manager also provides statistic information about
the state of the crawling process and basic access methods to the
collection.

\subsection{Using the GRID for Crawling and Feature Extraction}
\label{sec:extr}

We have considered GRID to be the right technology to obtain large
amount of computing power we needed. GRID is a very dynamic
environment that allows to transparently run a given application
on a large set of machines. In particular, we had the possibility
to access the EGEE (Enabling Grids for E-sciencE) European GRID
infrastructure\footnote{http://www.eu-egee.org/} provided to us by
the DILIGENT (Digital Library Infrastructure on Grid Enabled
Technology) IST project\footnote{http://www.diligentproject.org/}.

\begin{figure}
\centering
\includegraphics[width=10cm]{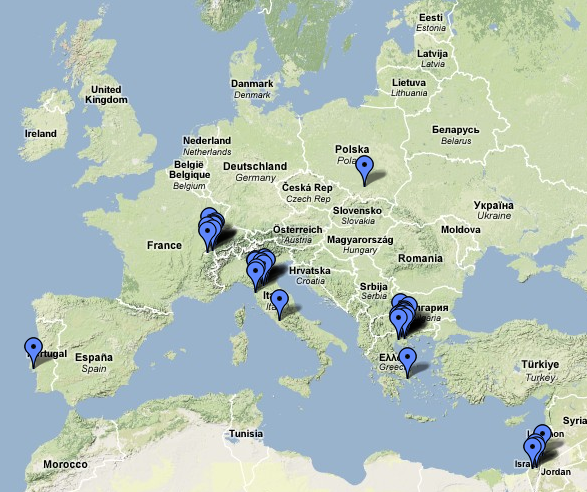}
\caption{\label{fig:geo}Machines collaborating on the crawling.}
\vspace*{-10pt}
\end{figure}

We were allowed to use $35$ machines spread across Europe (see
Figure~\ref{fig:geo}). We did not have an exclusive access to
these machines and they were not available all the time. Both
hardware and software configurations were heterogeneous: they had
various CPUs, memory, disk space, but also in the libraries,
software (e.g. Java), and Linux versions installed. Thus, we had
to build a self-contained crawling agent.

The crawling agent is logically divided into two modules.  The
first one accomplishes the communication with the image-id server,
crawls Flickr website, uses Flickr APIs, and sends the result of
the computation to the repository manager. This was coded in Java
to improve portability. However, since we could not assume the
presence of the Java virtual machine on every machine, we
incorporated into the crawling agents also a JVM and the required
Java libraries. Due to the latencies of the crawling task, the
crawling agent can instantiate a number of threads, each of them
taking care of processing a different image.
The settings which proved well is to have four threads per agent
(per one CPU core) and to process a maximum of 1,000 images. These
parameters induced computations times of 20 to 60 minutes
depending on the CPU speed.

The second module of the crawling agent is the MPEG-7 XM feature
extraction software. Since the MPEG-7 XM software is not maintained,
it has become incompatible with the recent compilers and
libraries versions. For this reason and for the heterogeneity of
the GRID, we encapsulated into the crawling-agents also all the
libraries it uses.

Submitting a job to a GRID infrastructure, the user does not have
a full control on the time and location where the job runs. The
GRID middleware software accepts the job description and schedules
it on the next available machine according to internal policies
related to the load of each node, the priority of the different
organization using the GRID infrastructure, etc. In our case, the
job always first downloads the crawling-agent package from our
repository-manager and then runs the software contained in the
package. The GRID provides a best-effort service, meaning that a
job submitted to the GRID may be rejected and never executed.
Indeed, there are several factors that may cause the failure of a
job submission. Out of the 66,440 jobs submitted, only 44,333 were
successfully executed that means that 33,3\,\% of the jobs failed
for GRID resources unavailability.

Our straightforward approach together with the self-scheduling of
images by each crawling agent has two important advantages. First,
in case the GRID middleware is not able to deploy the given job,
there would be no consequences in the remainder of the system,
especially, no image will be skipped. Second, in case of a
software update, it is just needed to replace the old version on
the repository manager with the new one.

Not all of the GRID machines were available through
the crawling period and, therefore, we also used a set of local
machines in Pisa which processed the images during the GRID idle
time. We thus reached the total of 73 machines
participating in the crawling and feature extraction process.



\begin{figure*}[t]
\begin{tabular}{cc}
\includegraphics[width=6cm]{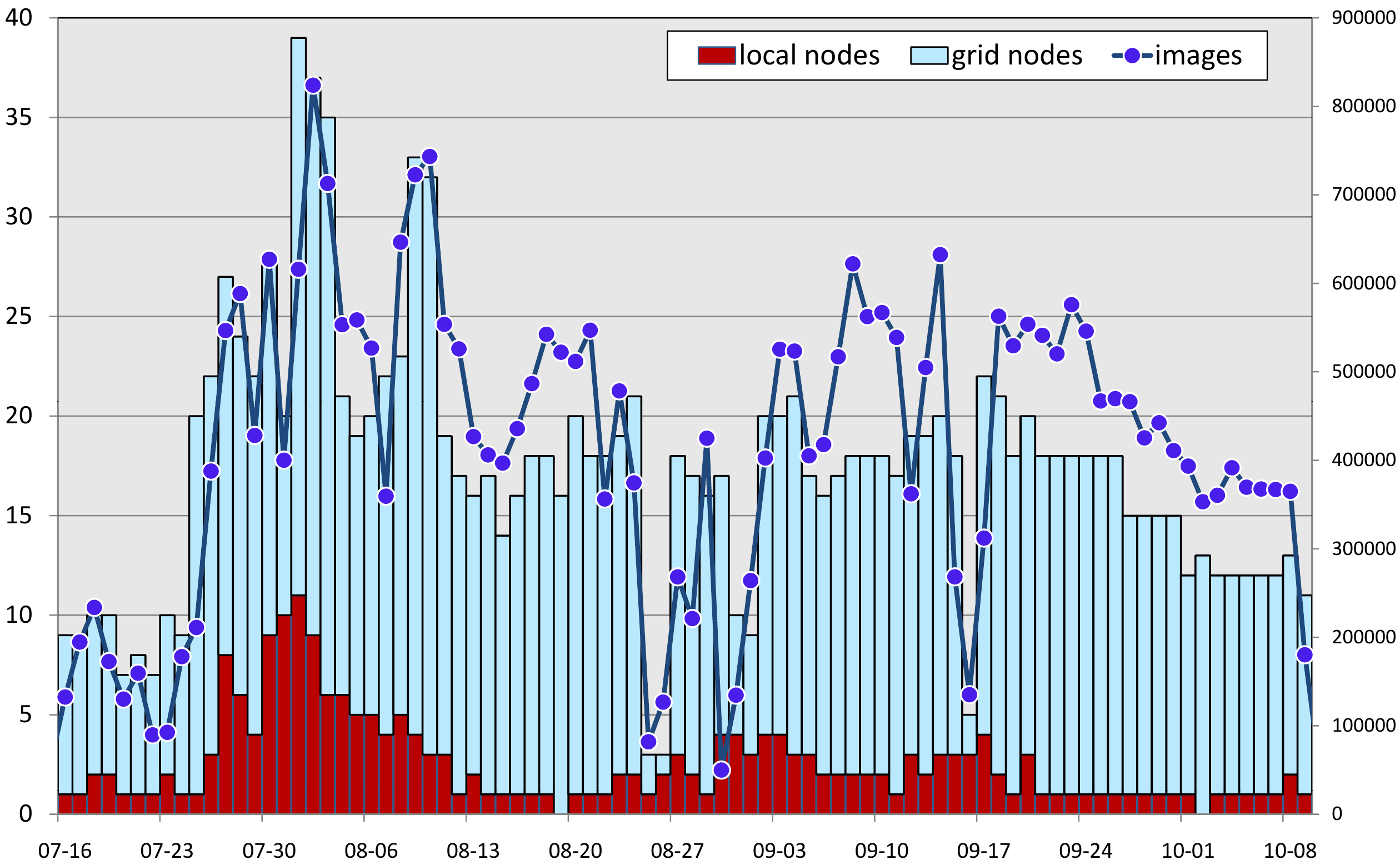}
\includegraphics[width=6cm]{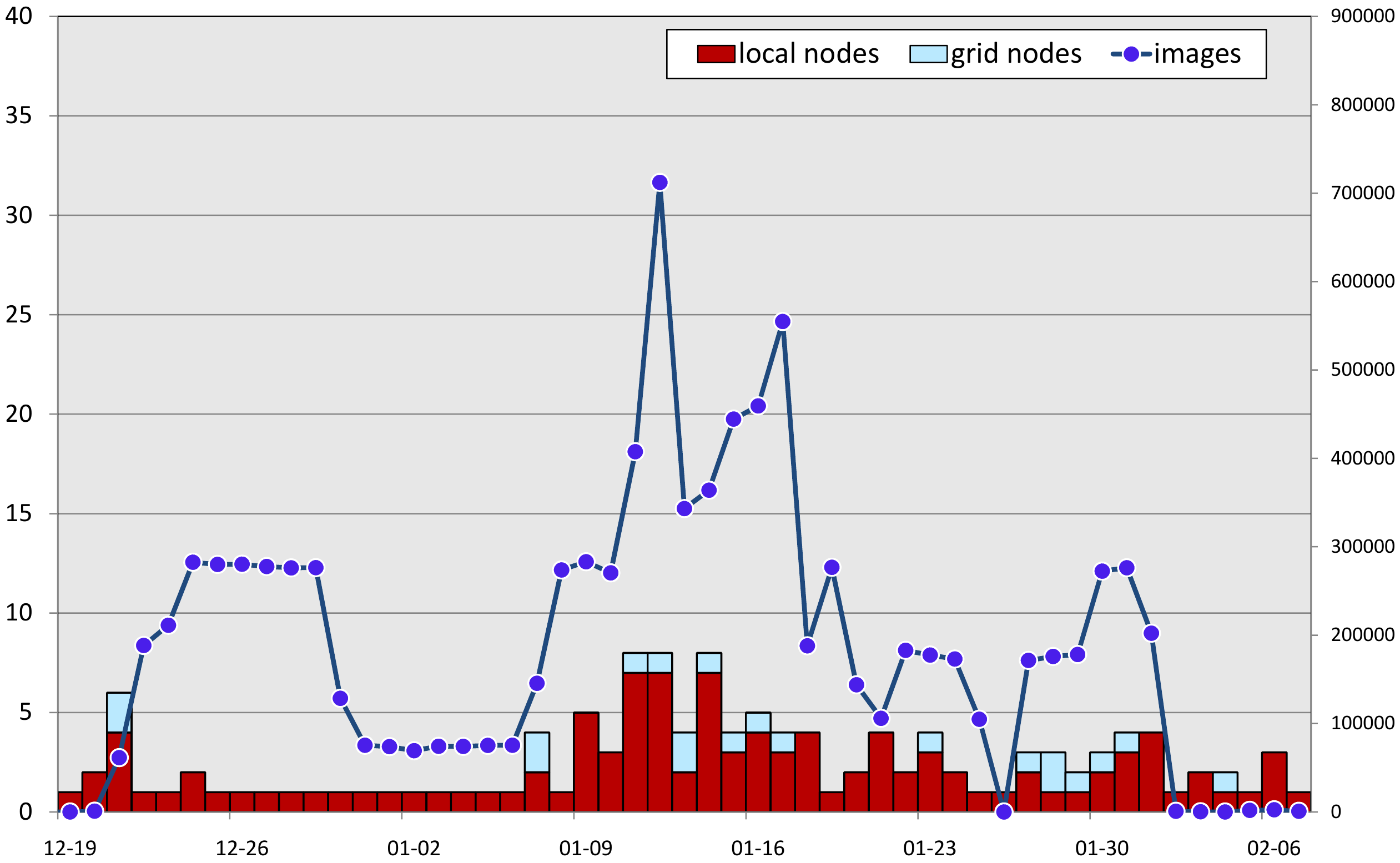}
\end{tabular}
\caption{\label{fig:gridslocal}Number of GRID and local machines
available during the two crawling periods:
from July 16$^{th}$ to October 9$^{th}$ 2007 (left) and
from December 19$^{th}$ 2007 to February 7$^{th}$ 2008 (right).}
\vspace*{-10pt}
\end{figure*}

The crawling process took place in two separate periods, both because
of GRID availability and because we needed to consolidate the data
after the first period. In Figure~\ref{fig:gridslocal}, we report
on the number of machines available during the crawling process.
During the first period, the GRID provided an average
of $14.7$ machines out of the 35 and, simultaneously, there were
2.5 local machines available, on average. Also the availability of
the machines during the day was unstable: The local machines
were mainly available over night while some of the GRID machines
were available only for a few hours per day. During the second
period, only one powerful multiprocessor machine was available
from the GRID, and we could continue the process only with our
local resources.

\begin{figure}
\centering
\includegraphics[width=10cm]{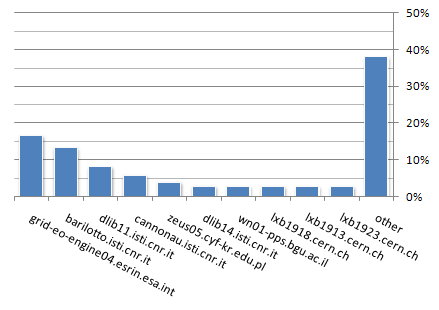}
\caption{\label{fig:chart}Number of images processed by each site.}
\vspace*{-10pt}
\end{figure}

Figure~\ref{fig:chart} reports the total number of images
processed by each site. The best machine (provided by the GRID)
processed about 17\,\% of the whole collection -- this is a very
powerful machine equipped with seven quad-core CPUs. The second
best is a local machine used only during the second phase is
equipped with two quad-cores Intel Xeon 2.0 GHz and it processed
about 13\,\% of the collection. These machines were the most
powerful and the most constantly available over time. However, the
largest total contribution came from a number of machines each of
which was able to process only a small number of images.



\subsection{The CoPhIR Collection}
\label{sec:cophir}

The result of the complex crawling and image processing activity described above is a test collection that served as the basis of the experiments on content-based image retrieval techniques and their scalability characteristics, in the context of the SAPIR project.

We reached, and passed, the target of 100 million images downloaded and processed at September 2008. 
Given the effort required in building a such large test collection, and the potential interest to the international research community, in order to make experiments in large-scale CBIR, we decided to make it available outside the SAPIR project scope.
The result is the CoPhIR (\textbf{Co}ntent-based \textbf{Ph}oto \textbf{I}mage \textbf{R}etrieval) Test Collection, managed by ISTI-CNR research institute in Pisa.

The data collected so far represents the world largest multimedia metadata collection available for research purposes, containing visual and textual information regarding 106 millions images\footnote{Actually $105,999,880$, due to 120 images that present XML data corruption, probably happened during the data gathering process from the Flickr website.}.

Each entry of the CoPhIR collection is an XML structure containing:
\begin{itemize}
\item identification information that allows to link and retrieve the corresponding image on the Flickr Web site;
\item the image textual data and metadata: author, title, description, GPS location, tags, comments, view count, etc.;
\item an XML sub-structure containing the information related to five standard MPEG-7 visual descriptors (see Section \ref{subsec:crawl}).
\end{itemize}


The disk space requirement for the CoPhIR collection constist of 245.3 GB for the XML data, 54.14 GB for the image content index, and 355.5 GB for the image thumbnails.

CoPhIR images come from 408,889 distinct authors, with a top contributor of 156,344 images (user \emph{conrado4}), and a \emph{median} value of images per author equal to 69.

The total number of comments in the collection is 55,188,775.
The total number of tag instances is 334,254,683, from a set of 4,666,256 distinct tags.
Table \ref{tab:tagFreq} shows the 30 most frequent tags, with their respective frequence.

\begin{table} 
\setlength{\tabcolsep}{1ex}
\centering
\begin{tabular}{|r|r||r|r||r|r|} \hline 
\multicolumn{1}{|l|}{Tag} & \multicolumn{1}{l||}{Frequency} & \multicolumn{1}{l|}{Tag} & \multicolumn{1}{l||}{Frequency} & \multicolumn{1}{l|}{Tag} & \multicolumn{1}{l|}{Frequency}\\ \hline
2006 & 2,950,783 & friends & 897,316 & summer & 623,637 \\ \hline
2007 & 2,073,932 & vacation & 895,217 & trip & 618,370 \\ \hline
wedding & 1,518,929 & beach & 806,749 & sanfrancisco & 605,606 \\ \hline
2005 & 1,473,134 & art & 747,901 & paris & 599,838 \\ \hline
party & 1,277,615 & nature & 728,711 & china & 595,492 \\ \hline
travel & 1,113,643 & nyc & 693,519 & usa & 591,824 \\ \hline
japan & 991,873 & birthday & 687,995 & water & 591,606 \\ \hline
family & 966,646 & italy & 661,737 & me & 573,267 \\ \hline
california & 924,140 & france & 657,811 & europe & 570,109 \\ \hline
london & 918,408 & music & 625,819 & flowers & 569,475 \\ \hline
\end{tabular}
\caption{\label{tab:tagFreq}The 30 most frequent tags in CoPhIR, with their respective frequencies.}
\end{table}

\begin{figure}[h!]
\centering
\includegraphics[width=6cm]{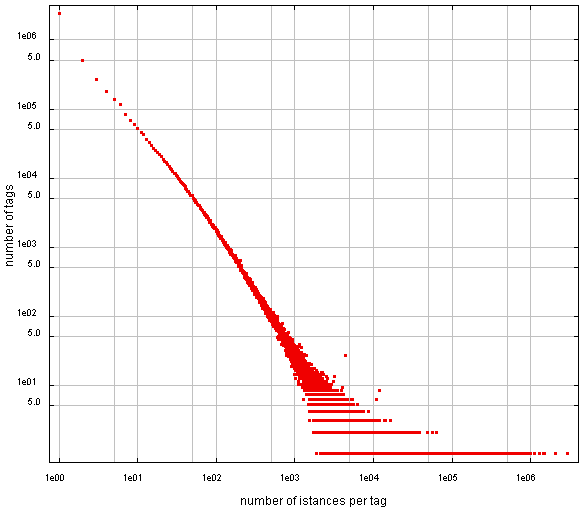}
\includegraphics[width=6cm]{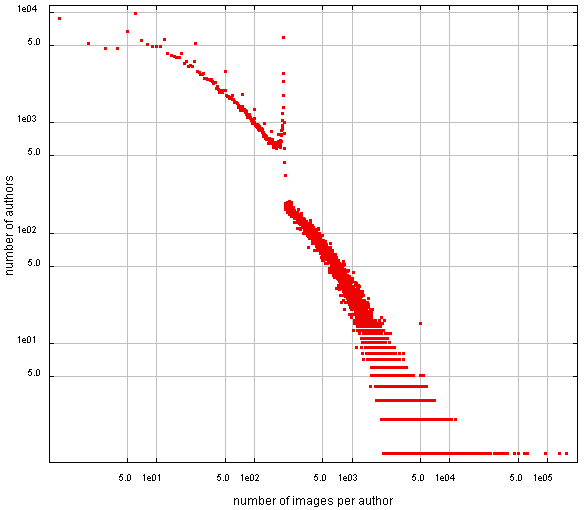}
\caption{\label{fig:TagOwnerDists}Frequency of tags with respect to the number of images they are associated to (left), and number owners with respect to number of images they have posted (right). The anomaly in the owners distribution is due to the fact that free accounts on Flickr have limitations, specifically only the most recent 200 images are made public.}
\end{figure}

Each image is thus associated on average with 0.52 comments and 5.02 tags.
However, the distribution of comments and tags among images is highly skewed, and follows a typical power law, as shown by the trend of graphs in Figure \ref{fig:TagOwnerDists}.

\begin{table}
\setlength{\tabcolsep}{1ex}
\centering
\begin{tabular}{|l|r|r|r|r|r||r|}
\hline
 & \multicolumn{5}{c||}{tag count} & \\ \hline
\multicolumn{1}{|c|}{comment count} &  \multicolumn{1}{l|}{0} & \multicolumn{1}{l|}{1-5} & \multicolumn{1}{l|}{6-20} & \multicolumn{1}{l|}{21-100} & \multicolumn{1}{l||}{$>$100} & \multicolumn{1}{l|}{Total} \\ \hline
0 & 31,753,762 & 45,546,507 & 15,276,035 & 438,756 & 6,778 & 93,021,838 \\ \hline
1-5 & 2,518,005 & 4,792,350 & 3,194,976 & 149,866 & 1,161 & 10,656,358 \\ \hline
6-20 & 302,894 & 675,849 & 882,819 & 66,278 & 205 & 1,928,045 \\ \hline
21-100 & 45,581 & 81,681 & 214,276 & 37,953 & 50 & 379,541 \\ \hline
$>$100 & 1,243 & 1,178 & 6,920 & 4,750 & 7 & 14,098 \\ \hline\hline
Total & 34,621,485 & 51,097,565 & 19,575,026 & 697,603 & 8,201 &\bf 105,999,880 \\ \hline
\end{tabular}
\caption{\label{tab:comstags} Distribution of images with respect to the number of comments and tags associated to them.}
\end{table}

Table \ref{tab:comstags} shows the number of images that have a specific number of comments/tags associated to them.
It turns out that the 87.76\% of the images have no comments, the 29.96\% of the images have no comments and no tags associated, while only the 14.16\% of the tagged images have at least one comment.
Only the 1.14\% of the images have at least six comments and six tags associated.

\begin{figure}[h!]
\centering
\includegraphics[width=12cm]{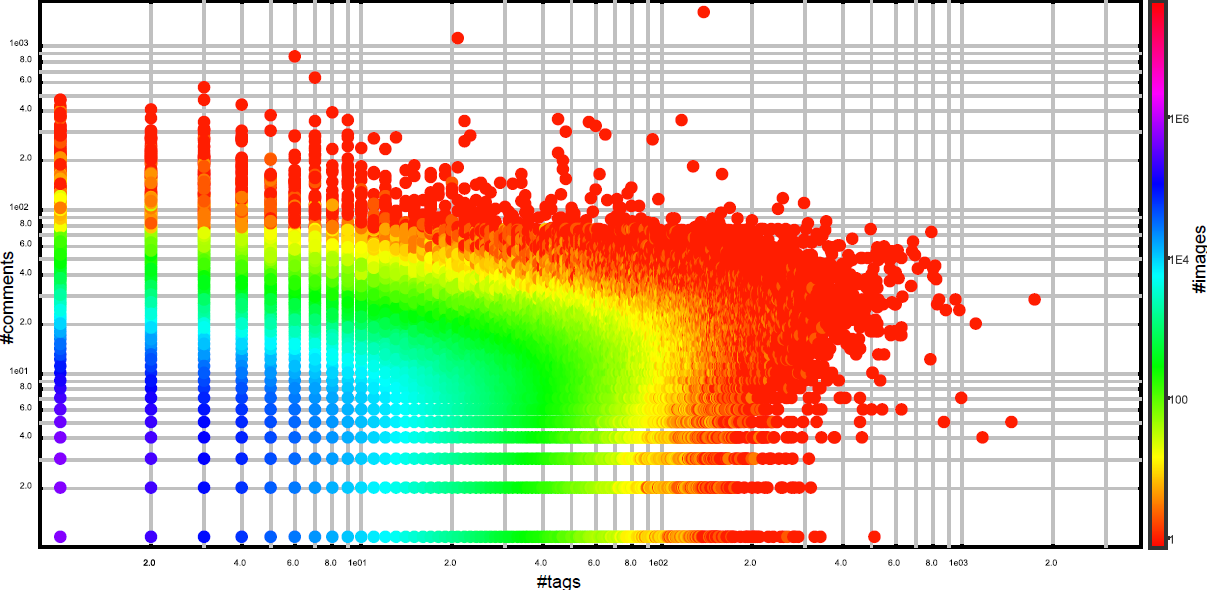}
\caption{\label{fig:comstags}Distribution of images with respect to the number of comments and tags associated to them.}
\end{figure}

Figure \ref{fig:comstags} shows a detailed plot of the distribution of images with respect to the number of comments and tags associated to them.

\begin{figure}[h!]
\centering
\includegraphics[width=10cm]{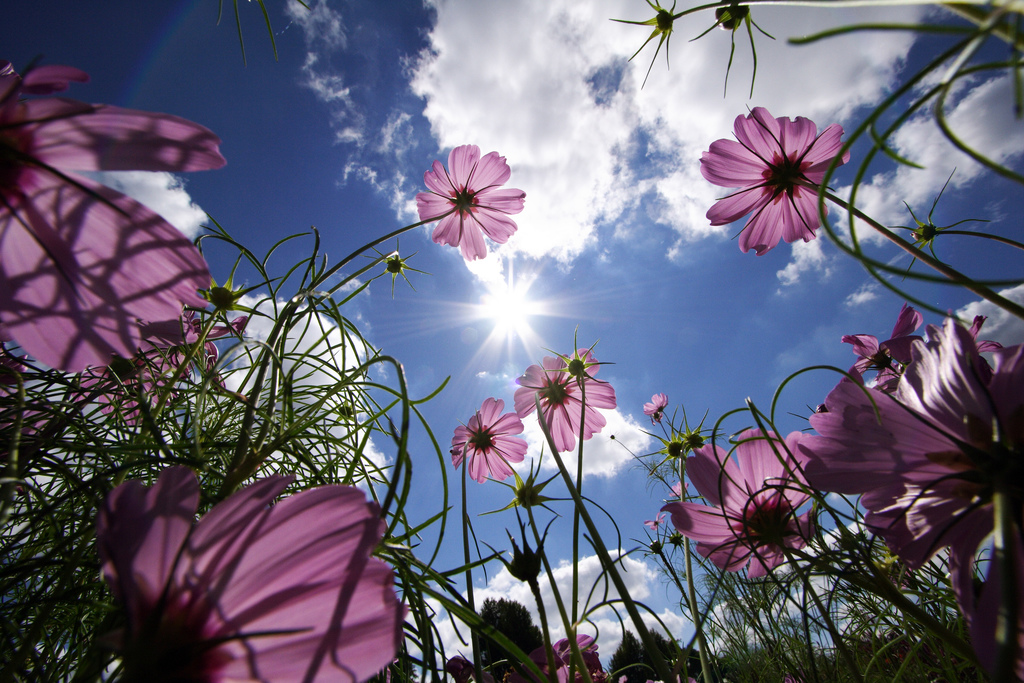}
\caption{\label{fig:mostViewsFavs}Most viewed and most favorited image in the CoPhIR collection.}
\end{figure}

In the collection, 66,532,213 images (62.77\% of the whole CoPhIR) have \emph{popularity} information, i.e., the count of the number of times an image has been viewed by any Flickr visitor, and the count of the number of users that have put the image in their \emph{favorite} image set.
The average number of views per image is 41.7, with a top value of 599,584 views, which correspond also to the most favorited image\footnote{Originally published at: \url{http://www.flickr.com/photos/alphageek/233472093/}\\ License info: \url{http://creativecommons.org/licenses/by-nc-sa/2.0/deed.en}}, selected by 3,662 users (see Figure \ref{fig:mostViewsFavs}).
Just half of the images with popularity information have registered more than 2 views (32,723,369 images, 49.18\%), and only 4,963,257 images (7.46\% of the part of CoPhIR with popularity information) have been marked has favorite by at least one Flickr user.

\begin{figure}
\centering
\includegraphics[width=12cm]{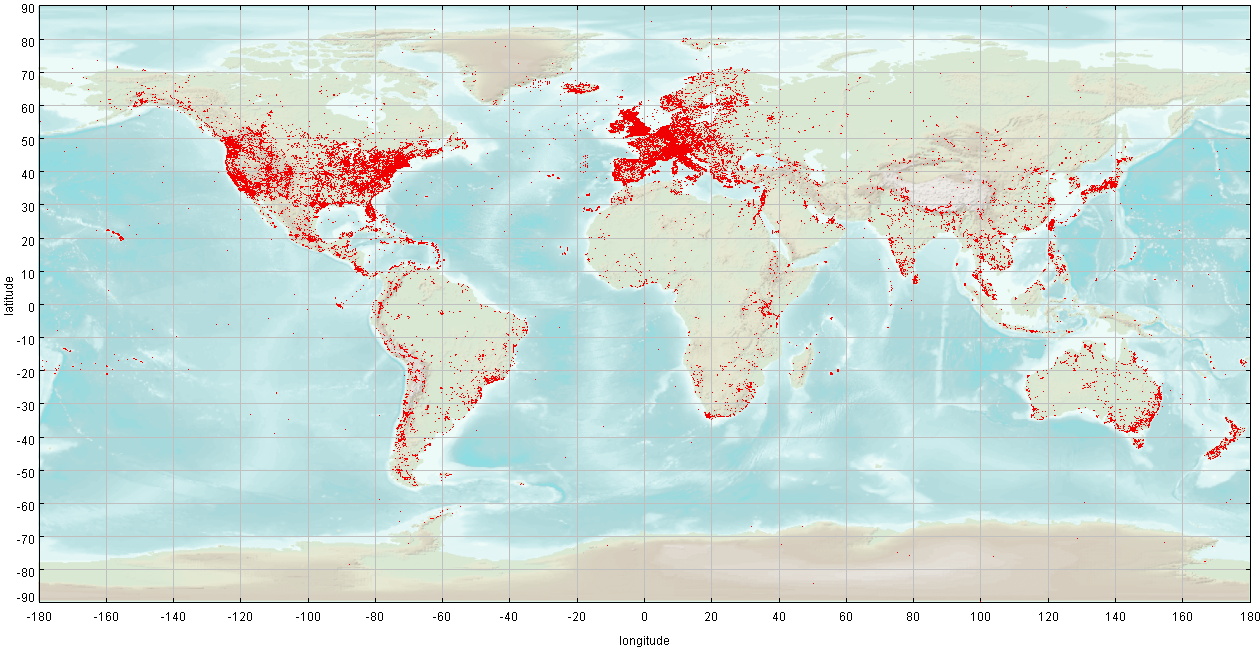}
\caption{\label{fig:geoloc}Map of the World's location referenced by CoPhIR images.}
\end{figure}

8,655,289 images (8.17\% of the whole CoPhIR) have \emph{geolocation} information associated to them. Figure \ref{fig:geoloc} shows a map of the referenced locations.



\subsection{Respecting the copyright constraints}

The scientific organizations (universities, research labs, etc.)  interested in experiments on CoPhIR have to register at the CoPhIR Web site\footnote{\url{http://cophir.isti.cnr.it/}}, and have to sign the CoPhIR Access Agreement establishing conditions and terms of use for the collection.

Such Access Agreement makes our use of the Flickr image content compliant to the most restrictive Creative Commons license.
Moreover, the CoPhiIR collection complies to the European Recommendation 29-2001 CE, based on WIPO (World Intellectual Property Organization) Copyright Treaty and Performances and Phonograms Treaty, and to the current Italian law 68-2003. 

Any experimental system built on the CoPhIR collection have to provide on the user interface, when displaying the Flickr images (or thumbnails) retrieved from the CoPhIR collection, an acknowledgment to the original image on the Flickr website, respecting all the rights reserved to the author of the such image.
The agreement states several conditions on the experimental applications, e.g., if original image is no more available on the Flickr website (deleted or made private) the corresponding entry should be removed from the indexed collection.



\section{Conclusions}
\label{sec:concl} No doubts that the scalability issue for new
digital data types is a real issue, which can be nicely
illustrated by difficulties with the management of the fast
growing digital image collections. In this paper, we focus on a
strictly related challenge of scalability: to obtain a non-trivial
collection of images with the corresponding descriptive features.

We have crawled a collection of over $100$ million high-quality
digital images, which is almost two orders of magnitude larger in
size than existing image databases used for content-base retrieval
and analysis. The images were taken from the Flickr photo-sharing
site which has the advantage of being a reliable long-term
repository of images enriched with a rich set of
additional metadata. Using a GRID technology, we have extracted
five descriptive features for each image. The features are defined
in MPEG-7 standard and express a visual essence of each image in
terms of colors, shape, and texture. This information is kept
handy in XML files -- one for each image -- together with the
metadata and links to original images in Flicker. This
unique collection is now open to the research community for
experiments and comparisons.  More than $50$ research institution worldwide already asked access to the CoPhIR collection by registering at the CoPhIR Web
site\footnote{http://cophir.isti.cnr.it}, and by signing the CoPhIR
Access Agreement, which establishes conditions and terms of use for the
collection.

\section*{Acknowledgments}
\noindent This work was partially supported by the IST FP6 Project SAPIR (Search In Audio Visual
Content Using Peer-to-Peer IR), contract no. 45128.

\bibliographystyle{splncs}
\bibliography{pisa}

\end{document}